# TRAJECTORIES OF LIGHT

Scientists have tried to elucidate the nature of light since the beginning of times. Newton, in the seventeenth century, asserted that light was formed by very tiny corpuscles. Later on, Huygens stated that light was a wave, basing this affirmation on its undulatory characteristics; however a wave has no significance without a medium that could transmit.

It was then when the concept of ether was invented, but Michelson and Morley`s experiments showed that ether does not exist. Simultaneously, they also demonstrated a more transcendental fact: The invariableness of the translational speed of light.

At the beginning of this century, Planck and Einstein did not approve the undulatory theory of light; rather supporting the corpuscle theory, calling them photons, each formed by one particle, traveling in a straight line when being gauged against dimensions larger than its wavelength. Nevertheless, when observing very small objects whose dimensions are in the same order of magnitude as the wavelength, then it appears as if light turns around; so its trajectory is curved and since it fits mathematically to sine equations, it has been believed that light trajectory is sinusoidal. However, in some way Newton found out that red light particles are larger than blue ones, being both corpuscles. He gave heed to the fact that when two light beams cross one another, practically no collisions occur. Only when beams travel in slightly convergent trajectories such as when experimenting interference of one and two slots, light interferes whit itself, canceling the beam if phase differential is 180 degrees, and doubling intensity if differential is 0 degrees or 360 degrees.

Electromagnetic radiations are not limited to visible light. Rather, the total spectrum from long-wave radio frequencies to cosmic rays is constituted by photons. Its behavior does not fit with the present theory, which considers photons as only one particle. Could somebody explain why photons have a spin equal to one, when most of subatomic particles have spins of + - 1/2? This sole fact suggests two particles rather than one. Or, as Hawking suggests, when particles have spins equal to one, a gyration of 360ª should be produced, so its position is the same than initial one.

A particle moving in a sinusoidal trajectory does not exist in nature, neither at microscopic nor at macroscopic scale. It goes against the most elementary laws of physics, especially against inertia.

Other common fallacy of our days, is to define light color by its wave length, because it is not constant, since it depends on its velocity, which in turn depends on the media in which it moves, such as air, vacuum, water, crystal, etc. If the equation stating that velocity is the product of wavelength by frequency, when velocity decreases, which other variable changes, frequency or wavelength? Everything seems to show that frequency remains constant while wavelength decreases in size, while color does not change when passing to a denser medium. Therefore, color depends only on frequency.

But how a photon, being a single particle, can "remember" its color? Or in other words, how a single particle can be frequency coded?

The present photon model cannot explain the different polarization, or how a photon moves in a lineal or elliptical polarization. If a photon was a single particle, how can it be explained that its total charge is zero, while it produces an electric and magnetic field?

The lack of coincidence of the present theory with the electromagnetic radiation performance has led me to the search of a theory that fits in with the duality particle-wave. The fact that photons follow the sinusoidal equations, suggest by itself that we are dealing with a single harmonic motion being its projection in a plane sinusoidal curve.

Single harmonic motion is characteristic of pendulums, springs, etc. and more universally of circular trajectories. It is characteristic of stars and atoms.

## PROPOSED PHOTON MODEL

Consider, in dynamic equilibrium, a photon formed by two particles whose electric charges are opposed, and for this reason they attract each other. This attraction is equilibrated by the centrifugal force generated when rotating one with the other. (See figure 1). By rotating electrical charged particles, a perpendicular magnetic field is generated, also perpendicular to a stationary observer.

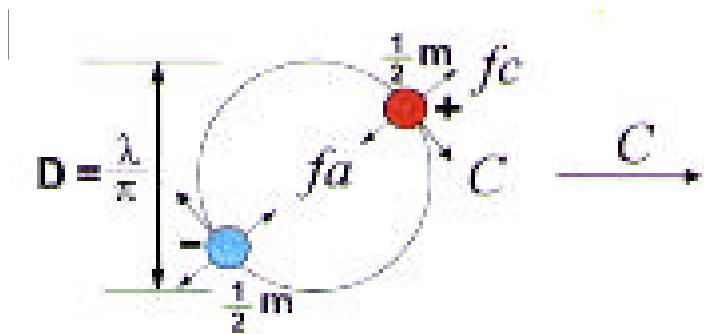

Charge

$$q = \sqrt{\frac{hc}{k}}$$

h = Planck Constant ($6.626 \times 10^{-34} \frac{joules}{sec.}$)

c = Light velocity in vacuum ($2.998 \times 10^8 \frac{m}{s}$)

k = Coulomb constant ($8.897 \times 10^9 \frac{Nm^2}{C^2}$)

Following are definitions of the considered photon:

**DISTANCE BETWEEN PARTICLES**: It is exactly the wavelength divided by $\rho$ (3.1416):

$$d = \frac{\lambda}{\rho}$$

**TANGENTIAL VELOCITY**: Coincides tangential velocity with the translational velocity, and in vacuum equals:

$$C = (2.998 \times 10^8 \frac{m}{s})$$

**PARTICLE MASS**: It is relativistic; in other words, depends on its frequency. Individual particle mass is equally spread. Otherwise, an unbalanced pattern in their trajectory will be obtained:

Individual particle mass: $\quad \frac{1}{2}m = \frac{hf}{2C^2}$

**PARTICLE CHARGE**: It is calculated equaling centrifugal force to the electrostatic attraction force as follows:

$$\frac{mC^2}{2r} = k\frac{q^2}{4r^2}$$

hence $\quad q^2 = 2\frac{mc^2 r}{k}$

Since $\quad m = \frac{hf}{C^2}$

$q^2 = \frac{2hfr}{k} \quad$ But $\quad C = f\lambda \text{ and } \lambda = 2\rho r$

Hence $\quad c = 2f\rho r \quad$ therefore $\quad fr = \frac{c}{2\rho}$

Then $\quad q^2 = \frac{hC}{\rho k} \quad$ therefore $\quad q = \sqrt{\frac{hC}{\rho k}}$

A constant value is obtained, by substituting the values:

$$q = \sqrt{\frac{(6.26 \times 10^{-34})(2.998 \times 10^8)}{3.1459(8.987 \times 10^9)}}$$

$$q = 2.653 \times 10^{-18} \, Coulombs$$

This charge if 16.558 times an electron charge.

**SYSTEM ENERGY**: It is the sum of translational kinetic energy plus rotational kinetic energy and is expressed:

Total energy: $\quad E_{total} = E_{translational-kinetic} + E_{rotacional-kinetic}$

$$E = E_{kt} + E_{kr} \quad (1)$$

$$E_{kt} = \tfrac{1}{2} \cdot \tfrac{1}{2} \cdot 2mc^2 = \frac{mc^2}{2} \quad (2)$$

$$E_{kr} = \tfrac{1}{2} \cdot \tfrac{1}{2} \cdot 2I w^2 = \frac{I w^2}{2} \quad (3)$$

The moment of inertia for a particle body is:

$$I = mr^2 \quad (4)$$

By substituting equation *(4)* into *(3)*:

$$E_{kr} = \frac{mr^2 w^2}{2}$$

And since the particle tangential velocity equals c, then:

$$r^2 w^2 = c^2 \quad \text{hence} \quad E_{kr} = \frac{mc^2}{2}$$

By substituting in *(1)*:

$$E = \frac{mc^2}{2} + \frac{mc^2}{2}$$

Which yields the well-known equation:

$$E = mc^2$$

This equation could not be obtained if the photon was a single particle.

**PATH**: It is variable, depending on polarization; if polarization is linear, the gyration plane coincides with the translational displacement, forming cycloidal trajectories.

When the translational movement is perpendicular to the rotation plane, then the trajectory is in spiral, called circular polarization. Any intermediate position will produce an elliptical polarization. (Fig.2).

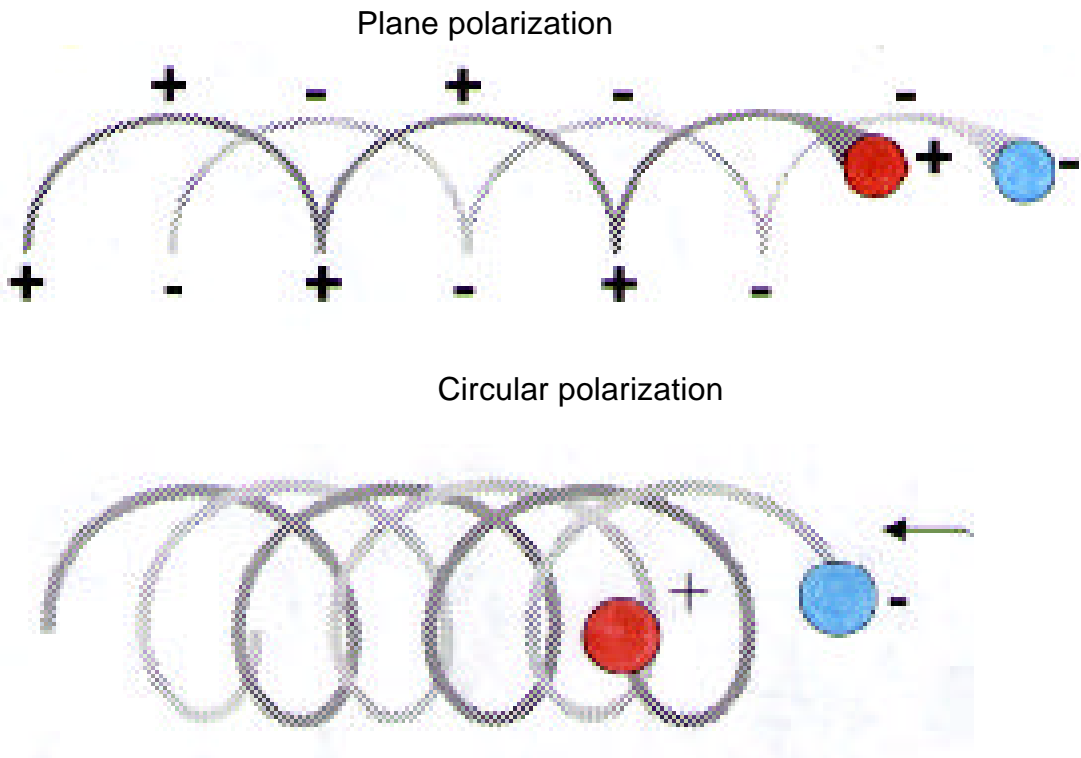

Plane polarization

Circular polarization

## TESTING PROPOSED MODEL

To test out if the above-described model fits the general performance, not only top visible light but to all electromagnetic spectrum, an experiment was conducted in the Electromagnetic Radiation Laboratory of the Engineering Faculty of the UNAM. Instruments employed were a Marconi oscillator (based on a Gunn diode), with 8 milliwatts power output, oscillating at 9,000 Mhz modulated with a 1 Khz square wave vertically polarized by means of a 1/4 wavelength antenna, reflected by a parabolic disk.

The wave produced was slightly intercepted by the edge of an aluminum screen, mounted on a ruler to measure horizontal displacements.

At the end of the wave's path, a receiving 1/4 wavelength antenna was mounted, adapted also to a horn antenna so as to minimize reflection and the consequent formation of standing waves. To the receiving antenna, a germanium detector was connected, and from this, to the vertical input of an oscilloscope to measure the intensity received by the receiving antenna. When the aluminum plate was displaced horizontally, a decrease in the radiation received was observed periodically, every 1.67 centimeters (this length is equivalent to 1/2 wavelength). Also the intensity received was exactly 1/2 of the amplitude of the square wave. If the wave path were sinusoidal, the minima peaks (vales) would be observed every full wavelength, and not every 1/2 wavelength. This is in agreement with the proposed hypothesis. However, the analysis of the model based on its performance would be:

A two-particle photon model has a total electrical charge of zero but forms an electrical field in the gyration plane, and a perpendicular magnetic field. This model explains why it is no possible to have a photon standing still since it is a system in dynamic equilibrium, by stopping it, this equilibrium is broken and the photon system is broken and the system is destroyed.

If all the degrees of freedom of two particles, rotational and translational moving system, all the polarization forms are exactly obtained, this is: Linear, circular and elliptical.

Particle mass depends on its frequency, and hence, it can be stated that it has a relativistic mass in direct proportion with its frequency. However, when standing still, its mass does not equals zero. It is a mass so small that cannot be measured with present day instruments. However, it is of the same order of magnitude (standing still) of the neutrino and the antineutrino mass.

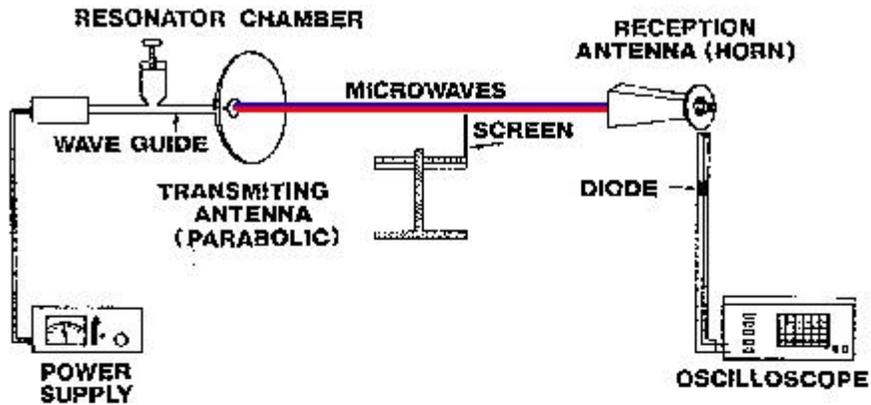

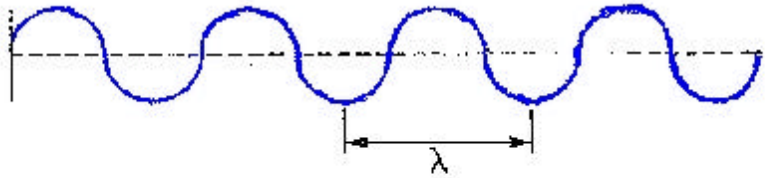

OLD SINUSOIDAL

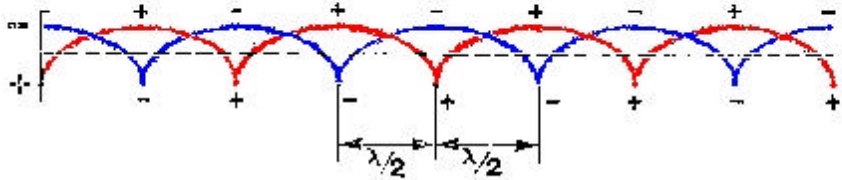

NEW HIPOTHESIS

## Second experiment

For this experiment, a simple L-C resonant oscillator, coupled to a resonant cavity which frequency was adjusted to 10.5 Ghz without modulation.

An horizontal movement system with a metallic plate was installed in the path of the microwave beam, making measures every millimeter, obtaining the following results:

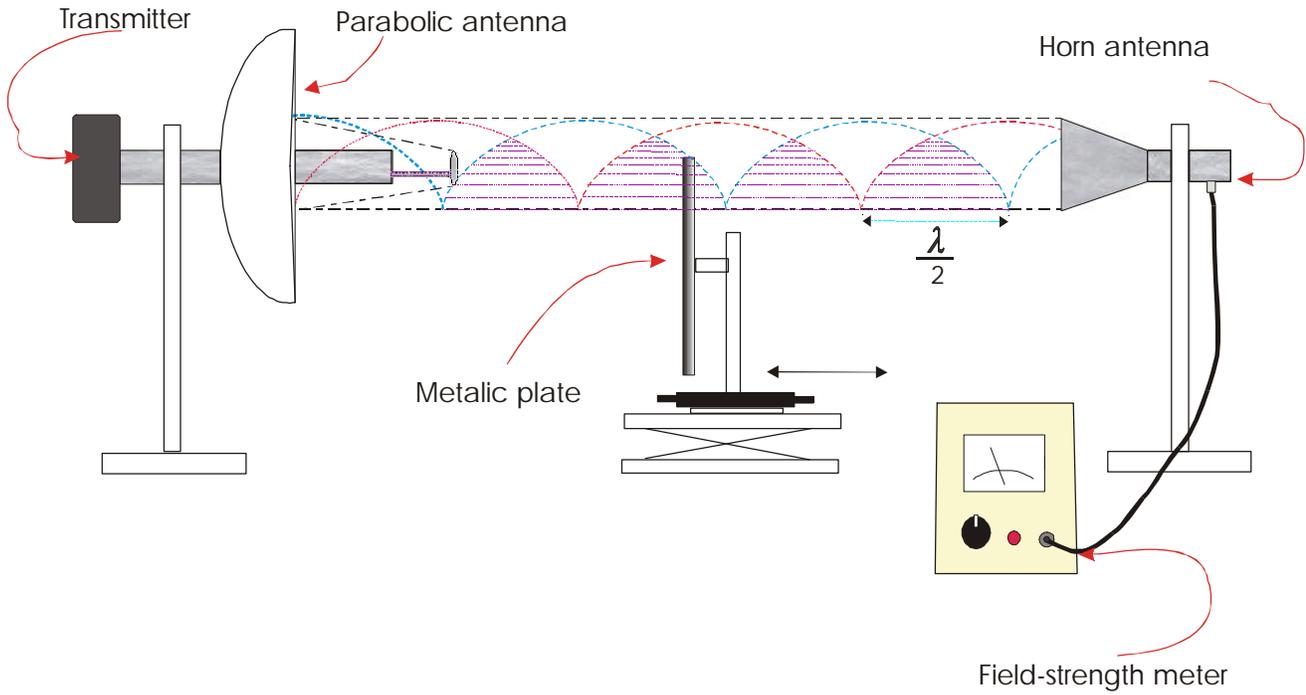

| Displacement in Cm | Field strength uA | Displacement in Cm | Field strength uA |
| --- | --- | --- | --- |
| 10 | 46 | 11.3 | 46 |
| 10.1 | 46 | 11.4 | 47 |
| 10.2 | 43 | 11.5 | 46 |
| 10.3 | 38 | 11.6 | 43 |
| 10.4 | 32 | 11.7 | 39 |
| 10.5 | 19 | 11.8 | 35 |
| 10.6 | 9 | 11.9 | 28 |
| 10.7 | 8 | 12 | 16 |
| 10.8 | 16 | 12.1 | 7 |
| 10.9 | 26 | 12.2 | 8 |
| 11.0 | 35 | 12.3 | 18 |
| 11.1 | 39 | 12.4 | 30 |
| 11.2 | 42 | 12.5 | 36 |
|  |  | 12.6 | 41 |

These results were graphicated where the horizontal axis coincides with the movement of the plate and vertically, the field strength was plotted.

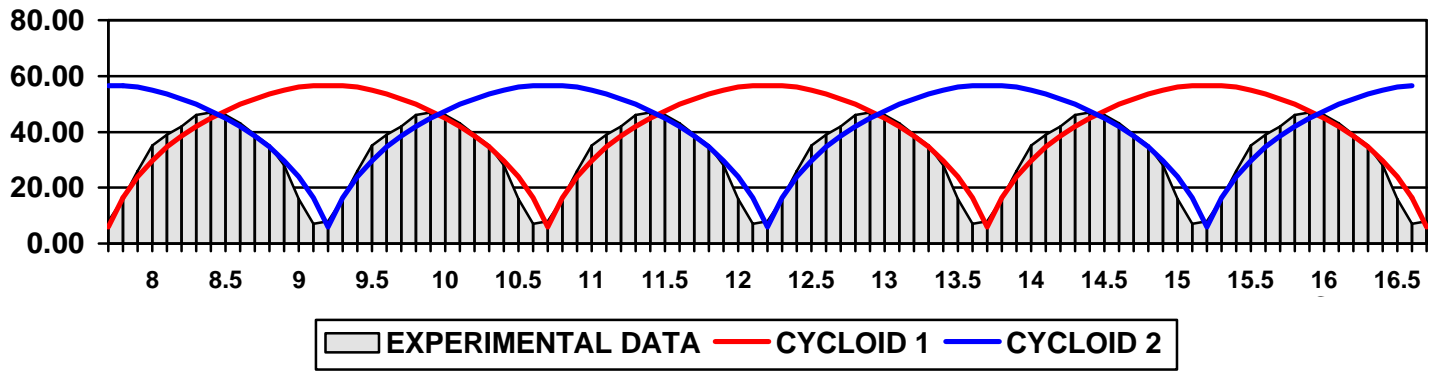

In this graphic it is clearly observed that the trajectories are not senoidals but coincides exactly with the empty spaces of the plate following the cycloidal paths, which are generated by a spinning wheel, which the tangential speed is the same as its translational speed.

An experiment is yet to be made using the same equipment in which helical antennas will produce a circular polarization in order to obtain the field intensity pattern.

It is demonstrated that the path of the electromagnetic radiations is not senoidal. Instead, the trajectories are cycloidal in plane polarization. The simple harmonic movement generates this path.

This equipment is very simple and not expensive, it is the same used by the universities teaching microwave antennas and communications. Therefore we encourage you to do the same experiment.

It is very interesting to note that when generating electromagnetic radiations, an electron always takes part, which suggests that electrons are not simple particles.

By means of the proposed model, the change of the trajectory angle is explained, when waves are refracted by passing from a less denser to a more denser media and vice versa. One of the particles is slowed down by electrostatic attraction, but the other, follows with the same velocity for a very short period of time, but enough to change the angle of the trajectory of the system.

The corpuscle properties of electromagnetic radiations are set by the existence of the two particles but, on the other hand, the undulating properties are produced by the rotation of these, solving the dual personality problem of corpuscle-wave. This model is compatible with the way the system frequency is coded, as compared with the way the time is set when the earth makes a full turn around the sun.

It has been said that light is its own antiparticle, or in other words, it is not possible to determine if observed light comes from matter or antimatter, opening the possibility that light may have two electric charges, since one of the differences between matter an antimatter consist in the sign of their electric charges, when they have them. The particles that form the photon, for their extremely small mass when standing still, could be called positrino and electrino.

The fact that Newton did not observe collisions in two light beams that cross one another is due to the fact that photons do not have matter between their particles, and hence, the probability of a collision is greatly reduced due to the large amount of hollow space.

# ING. VICTOR M. URBINA BOLLAND

Av. Urbina 4, Parque Industrial
Naucalpan Edo. de México
México
CP 53489
Tel (525) 308-2315
Fax (525) 301-0863
vurbina@qfoliar.com.mx

**GLOSSARY**

m Photon mass
c light velocity in vacuum ($2.998 \times 10^8 \frac{m}{s}$)
r particle radius of gyration
k Coulomb constant ($8.897 \times 10^9 \frac{Nm^2}{C^2}$)
f frequency, Hz.
$l$ wavelength, m.
h Planck constant ($6.626 \times 10^{-34} \frac{joules}{sec.}$)
$p$ 3.1416
q particle charge ($2.653 \times 10^{-18} Coulombs$)
E energy in Joules.
d distance between particles (2r)
Ekt translational kinetic energy, *Joules.*
Ekr rotational kinetic energy, *Joules.*
I inertia moment $Kg \cdot m^2$
$w$ angular velocity, rad / sec.